\documentclass[aps,pre,11pt,superscriptaddress,floatfix,tightenlines,showpacs,longbibliography,notitlepage]{revtex4-1}
\usepackage{graphicx}
\usepackage{bm}
\usepackage{amsfonts}
\usepackage{color}
\usepackage{amsmath}    
\usepackage{epsfig}
\usepackage{soul}
\usepackage{hyperref}   

\newcommand{\bu}{\boldsymbol{u}}

\newcommand{\bx}{\boldsymbol{x}}

\newcommand{\bk}{\boldsymbol{k}}

\newcommand{\ictsaddress}{International Centre for Theoretical Sciences, Tata Institute of Fundamental Research, Bangalore 560089, India}
\newcommand{\iitsaddress}{Department of Chemical Engineering, Indian Institute of Technology Bombay,
Mumbai 400076, India}
\newcommand{\nordita}{Nordita, KTH Royal Institute of Technology and Stockholm University, Roslagstullsbacken 23, 10691 Stockholm, Sweden}

\begin{document}
\title{Lagrangian Irreversibility and Eulerian Dissipation in Fully-Developed Turbulence}
\author{Jason R. Picardo}
\email{jrpicardo@che.iitb.ac.in}
\affiliation{\iitsaddress}
\author{Akshay Bhatnagar}
\email{akshayphy@gmail.com}
\affiliation{\nordita}
\author{Samriddhi Sankar Ray}
\email{samriddhisankarray@gmail.com}
\affiliation{\ictsaddress}
\begin{abstract}

We revisit the issue of Lagrangian irreversibility in the context of recent
results [Xu, \textit{et al.}, PNAS, \textbf{111}, 7558 (2014)] on
flight-crash events in turbulent flows and show how extreme events in
the Eulerian dissipation statistics are
related to the statistics of power-fluctuations for
tracer-trajectories. Surprisingly, we find that particle trajectories in intense dissipation zones
are dominated by energy gains sharper
than energy losses, contrary to flight-crashes, through a pressure-gradient driven \textit{take-off} phenomenon.
Our conclusions are rationalised by analysing data from simulations of three-dimensional intermittent turbulence, as well as from non-intermittent 
decimated flows. Lagrangian irreversibility is found to persist even in the latter case, wherein fluctuations of the dissipation rate are shown to be relatively mild and to follow probability distribution 
functions with exponential tails. 

\end{abstract}

\maketitle

The significant advances in Lagrangian techniques, especially in experiments,
over the last couple of decades, have allowed us to revisit some of the
more fundamental aspects of fully developed, statistically homogeneous, isotropic
three-dimensional turbulence~\cite{Yeung-Rev}. These include the ideas of irreversibility and
intermittency which form the two cornerstones for an Eulerian description of
such flows. Indeed, intermittency effects, which ensure that the Kolmogorov
description for turbulence is not \textit{exact}~\cite{Frisch-CUP}, show up often more strongly
in Lagrangian measurements suggesting an equivalence between these two
frameworks. This equivalence, borne out through bridge
relations which relate the scaling exponents of velocity structure
functions evaluated in one framework to the other~\cite{Borgas,Luca-PRL-2004,Mitra2004,Schmitt}, 
remains a much studied problem even now~\cite{Homann2009,Bentkamp2019}.

Much more recently, an important development came by way of using Lagrangian
probes to measure, and understand, time-irreversibility in turbulent flows: the kinetic energy of fluid particles was found to fluctuate with a marked temporal asymmetry---\textit{flight-crash} events---of gradual energy gains interspersed with sudden, rapid losses~\citep{Xu-PNAS,flight-crash-vortex,Bhatnagar-Flight-Crash,Maity}.
From an Eulerian perspective, the irreversibility of turbulent flows follows directly from the fact
that such flows, or solutions to the equations which model such flows, are
dissipative~\citep{Verma2019} with a non-vanishing energy flux. The more striking aspect of this work~\cite{Xu-PNAS} is how the
irreversibility of the flow manifests itself in a Lagrangian
framework, giving rise to the notion of Lagrangian irreversibility.

\begin{figure*}
\includegraphics[width=\textwidth]{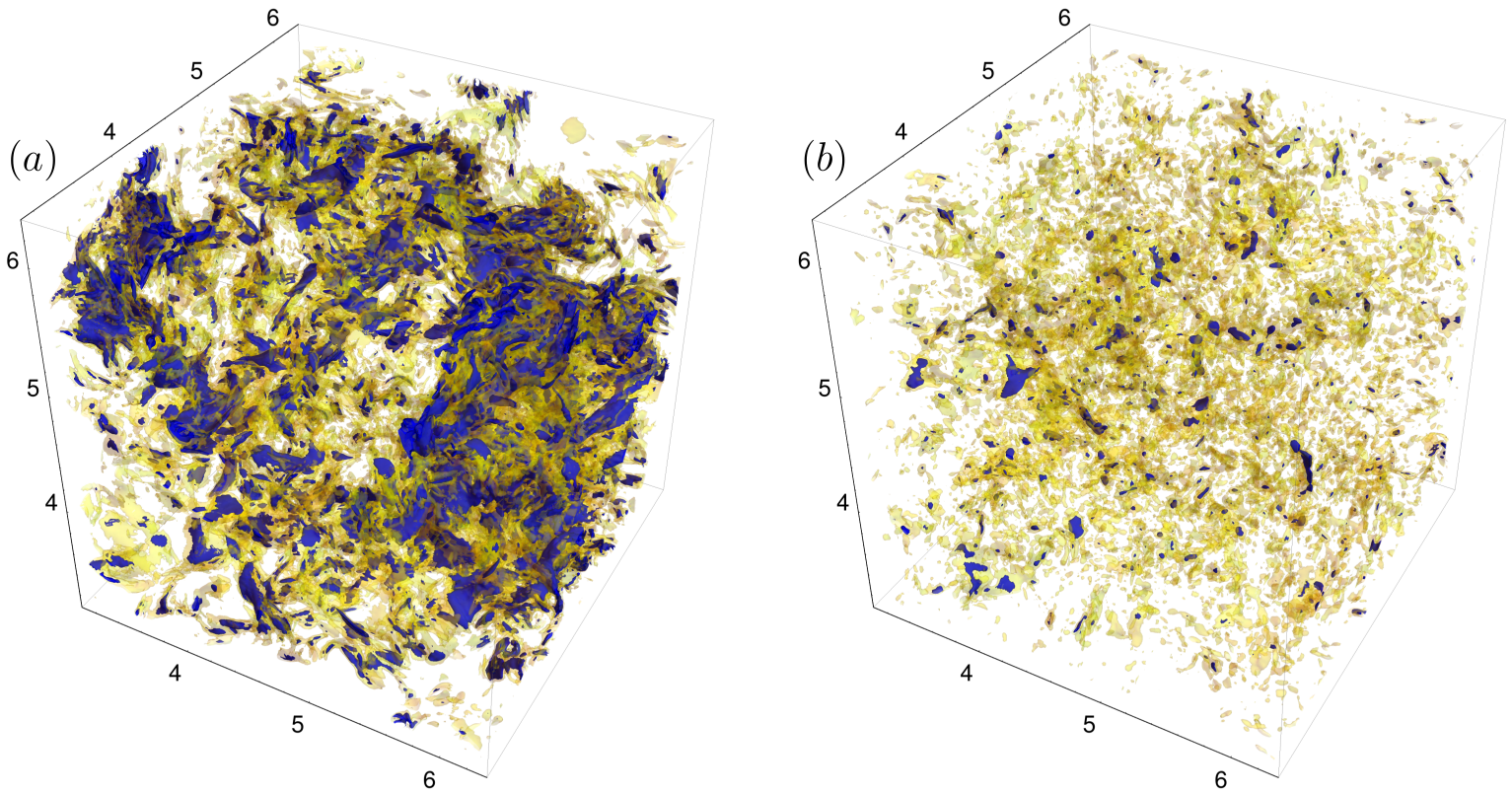}
	\caption{Contours of intense Eulerian energy dissipation ($\varepsilon = 6\bar \varepsilon$ in blue and $\varepsilon = 4 \bar \varepsilon$, in yellow) from snaphots of (a) the full 
	 3D flow and (b) a homogeneously decimated field ($\alpha = 0.1$). Fractal decimation results 
	 in a similar calming of the $\varepsilon$ field.}
\label{Fig:eps}
\end{figure*}

In this paper, we revisit the flight-crash phenomenon and critically examine if this Lagrangian measure
of irreversibility is related in any way to extreme, small-scale fluctuations of the Eulerian dissipation field. Indeed, it is tempting to associate the energy \textit{crashes} of a fluid particle with its passage through the sheet-like intense dissipation zones that proliferate fully-developed, intermittent turbulence [cf. Fig.~\ref{Fig:eps}(a)]. Contrary to this expectation, 
we show that the
strongest flight-crashes, which lead to a finite measure of irreversibility,
come from regions of the flow which are quiescent. Moreover, we explain why fluid particles in intense dissipation zones do not crash, but gain energy rapidly and \textit{take-off}.

We substantiate the association, or lack thereof, between
small-scale intense dissipative regions and irreversibility by performing additional
calculations on a system---the decimated Navier-Stokes equation~\cite{Decimated-PRL-2012,Ray-Review}---which
mimics statistically, homogeneous isotropic turbulence without
intermittency~\cite{Luca-PRL-2015,Buzzicotti-PRE,Buzzicotti-NJP,Ray-PRF2018}. We find that even in the limiting case of
\textit{non-intermittent turbulence}, wherein extreme small-scale structures are suppressed, Lagrangian irreversibility, as measured
through flight-crashes, persists.  We therefore show, via a careful
measurement and analysis of data from both three-dimensional intermittent and
decimated non-intermittent turbulence, that Lagrangian irreversibility is \textit{not} rooted in the extreme small-scale dissipative
structures of the flow, and that their relationship is neither intuitive, nor straightforward.

Our investigations are based, in part, on the three-dimensional (3D) incompressible
Navier-Stokes equations, solved numerically on a $2\pi$-periodic cubic box,
through a standard pseudo-spectral method with a second-order Adams-Bashforth
scheme for time-marching, to yield the fluid velocity field ${\bf u}$. We use $N=512^3$ collocation points and a constant energy-injection 
scheme to drive our system to a statistically steady state characterised by a 
Taylor-scale Reynolds number $Re_\lambda = 110$. Once our flow reaches this steady state, 
we seed, randomly, the flow with $10^5$ Lagrangian (tracer) non-interacting particles. The dynamics in phase-space
of each of these Lagrangian particles is determined by their position and velocity ${\bf v}_p = {\bf u}({\bf x}_p)$, 
where ${\bf u}({\bf x}_p)$ is the fluid velocity at the particle position ${\bf x}_p$. Given that we solve for the Eulerian 
fluid velocity on a regular cubic grid and that typically particle positions are off-grid, we resort 
to a tri-linear interpolation scheme to obtain ${\bf u}$ at the particle position ${\bf x}_p$; we have checked 
the accuracy of our scheme by benchmarking our Lagrangian statistics with results reported earlier from 
several other groups.

We also look at Lagrangian dynamics in a different class of turbulent flows which are obtained as solutions ${\bf v}$ of 
the incompressible \textit{decimated} Navier-Stokes equation (NSE)~\cite{Decimated-PRL-2012}. 
The decimated Navier-Stokes 
equation is obtained from the 3D equation by using a generalised Galerkin-projector $\cal P$:
\begin{equation}
{\bf v}(\bx,t)= {\cal P} \, {\bf u}(\bx,t)=\hspace{-1mm}
\sum_{{\bk}}\hspace{-1mm} e^{i {\bk \cdot \bx}}\,\gamma_{\bk}\hat{\bu}(\bk,t).
\label{eq:decimOper}
\end{equation}
The parameters  
\begin{equation}
\gamma_{\bk} = \begin{cases}1 ~ \text{with probability}~h_k\\0
  ~\text{with probability}~1-h_k \ , \quad k\equiv |\bk| 
 \end{cases}
\end{equation}
allow us to eliminate a random---but frozen in time---subset of Fourier modes leading to the evolution of the decimated velocity 
field, via 
\begin{equation}
\partial_t {\bf v} = {\cal P}[- {\bf \nabla}P -
(\bf v \cdot {\bf \nabla}  \bf v )]  + 
  \nu \,\nabla^2 {\bf v} +  {\bf F}\,. 
\label{eq:decimNS}
\end{equation}

This surgical removal of a pre-chosen set of Fourier modes, at all times, by
using the generalised Galerkin projector not only on the quadratic term but also on
the initial conditions and the forcing ${\bf F}$, leads to the evolution
of the decimated velocity field on a fractured Fourier lattice. 
The nature of the fracturing of the Fourier lattice depends on the
way in which $h_k$ is chosen. One possibility is $h_k \propto (k/k_0)^{D-d}$,
with $0< D \le d$ (where $k_0$, a reference wavenumber, is conveniently set to
1), which leads to a fractal Fourier grid with a bias towards the removal of Fourier
modes with larger values of $k$.  Such an approach~\cite{Decimated-PRL-2012}---{\it fractal
decimation}---has the advantage of allowing an easy interpretation of the
resulting dynamics in terms of a dimension $D$, corresponding to the fractal
dimension of the Fourier lattice (embedded in a $d$-dimensional space), in a way 
different from earlier methods~\cite{Fournier-Frisch,Dario}. Therefore, it allows us 
to obtain equilibrium solutions~\cite{Lvov-PRL}, and has led to several studies at the interface of turbulence 
and equilibrium statistical physics~\cite{Inverse-PRL,Luca-Helical,Banerjee-PRE,Decimated-Shear-PRE,Tom-EPL,Divya}. An alternative, unbiased protocol that avoids the preferential removal of small-scales is
{\it homogeneous decimation}~\cite{Buzzicotti-NJP,Ray-PRF2018}, $h_k = 1-\alpha$
($0 \le \alpha \le 1$), which ensures that the probability of removal of a Fourier
mode is independent of $k$.  In this study, we use both fractally and
homogeneously decimated turbulent
velocity fields, along with the non-decimated, fully three-dimensional turbulent flow. Lagrangian trajectories are tracked in the
decimated flow field~\cite{Buzzicotti-NJP} in ways exactly similar to that in the full
three-dimensional flow, with ${\bf v}_p = {\bf v}({\bf x}_p)$.  

Following these Lagrangian trajectories, we measure the evolution of the kinetic energy ${E = ({\bf v}_p \cdot {\bf v}_p)/2}$ and calculate the power $p = \frac{dE}{dt}$, whose fluctuations bear the imprint of Lagrangian irreversibility. As shown in the pioneering work of Xu, \textit{et al.}~\cite{Xu-PNAS}---later
extended by Bhatnagar, \textit{et al.}~\cite{Bhatnagar-Flight-Crash}---the distribution of $p$ for time-irreversible trajectories is negatively skewed, indicative of relatively rapid energy losses.  

\begin{figure}
\includegraphics[width=0.85\columnwidth]{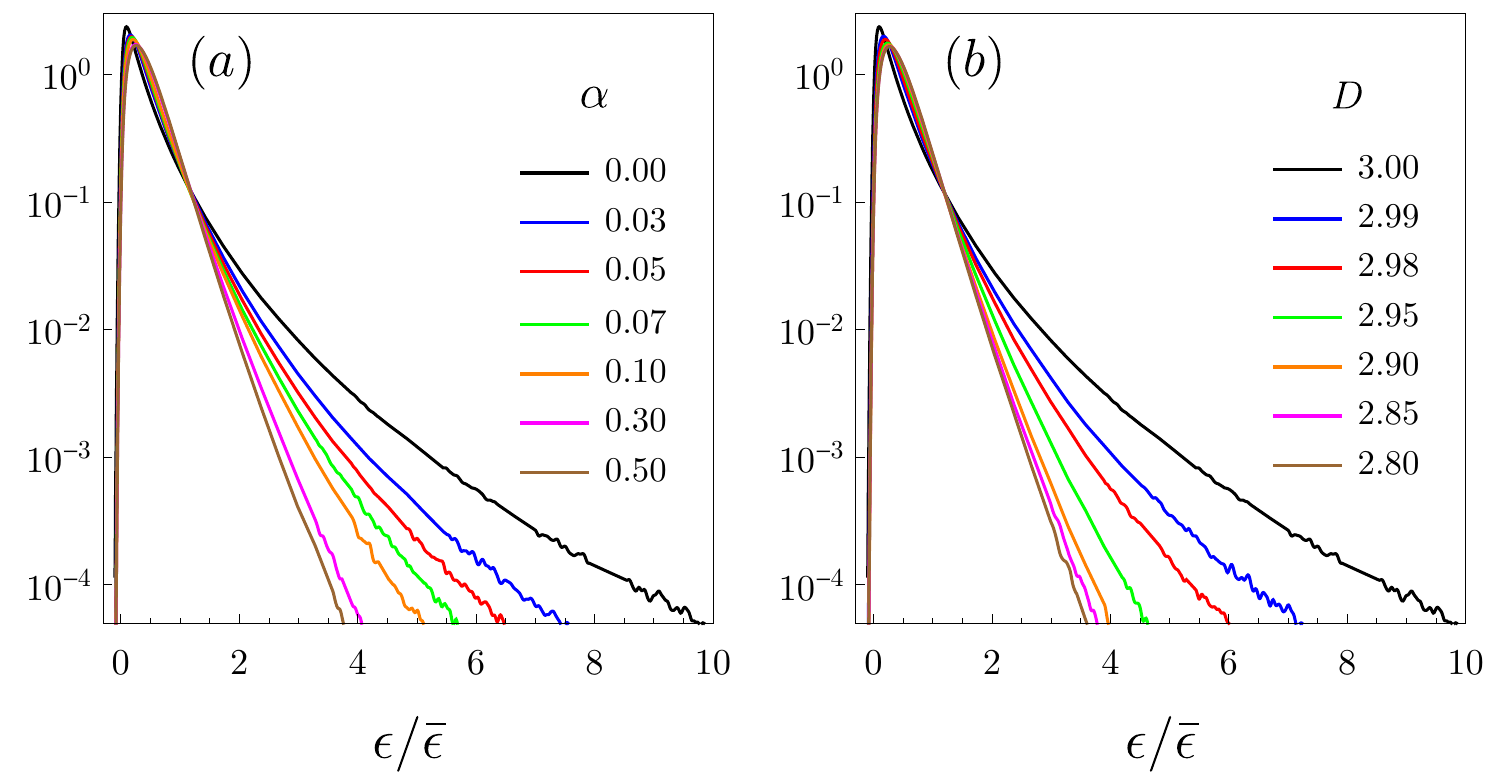}
	\caption{Probability distribution function (pdf) of the energy dissipation rate in the (a) homogeneously and 
	(b) fractally decimated Navier-Stokes equation for different values of $\alpha$ and $D$, as well as for the 3D flow ($\alpha = 0$; 
	$D = 3$). The tails of the pdf change from an approximately 
	log-normal to exponential distributions as the flow is increasingly decimated.}
\label{Fig:eps-pdf}
\end{figure}

We now turn to the key question motivating our study: Is there a direct causal connection of intense
Eulerian dissipative structures to flight-crashes and, hence, Lagrangian
irreversibility?  In Fig.~\ref{Fig:eps}(a), we show a snapshot, from our 
3D simulations, of contours of intense Eulerian
dissipation $\varepsilon = 2 \nu S_{ij} S_{ij}$, where ${ S} = (\nabla {\bf u}+\nabla {\bf u}^T)/2$ \citep{Frisch-CUP}. Clearly, the regions of extreme dissipation, though
inhomogeneously distributed and intermittent, are not rare, even for the large threshold of $6\bar \varepsilon$ (blue regions). Hence, a typical
Lagrangian trajectory would encounter such regions with a finite frequency, 
suggesting the plausible scenario of extreme dissipation zones serving as sinks
in which tracers lose energy rapidly. If true, this
would imply that flight-crashes, as a signature for Lagrangian
irreversibility, must be pegged to the statistics of the extreme
events underlying Eulerian dissipation in 3D turbulence. In testing this conjecture, a
decimated turbulent flow field is a useful setting. This is because, as shown in Fig.~\ref{Fig:eps}(b) (for
$\alpha = 0.1$), the dissipation field for decimated turbulence is much more uniform with fewer extreme events. Indeed the
probability distribution function (pdf) of $\varepsilon$, which is empirically
known to be close to log-normal for 3D turbulence~\cite{Frisch-CUP}, shows an increasingly exponential behaviour with the
reduction of the effective degrees of freedom through decimation, as shown in
Fig.~\ref{Fig:eps-pdf}. [Note: This suppression of extreme dissipation events by decimation coincides with the loss of small-scale intermittency of the velocity field, as evidenced by, e.g., the kurtosis of the longitudinal velocity increment approaching a Gaussian value of 3 with increasing decimation (see Fig. 4 of ref.~\cite{Luca-PRL-2015}).]

\begin{figure}
\includegraphics[width=0.55\columnwidth]{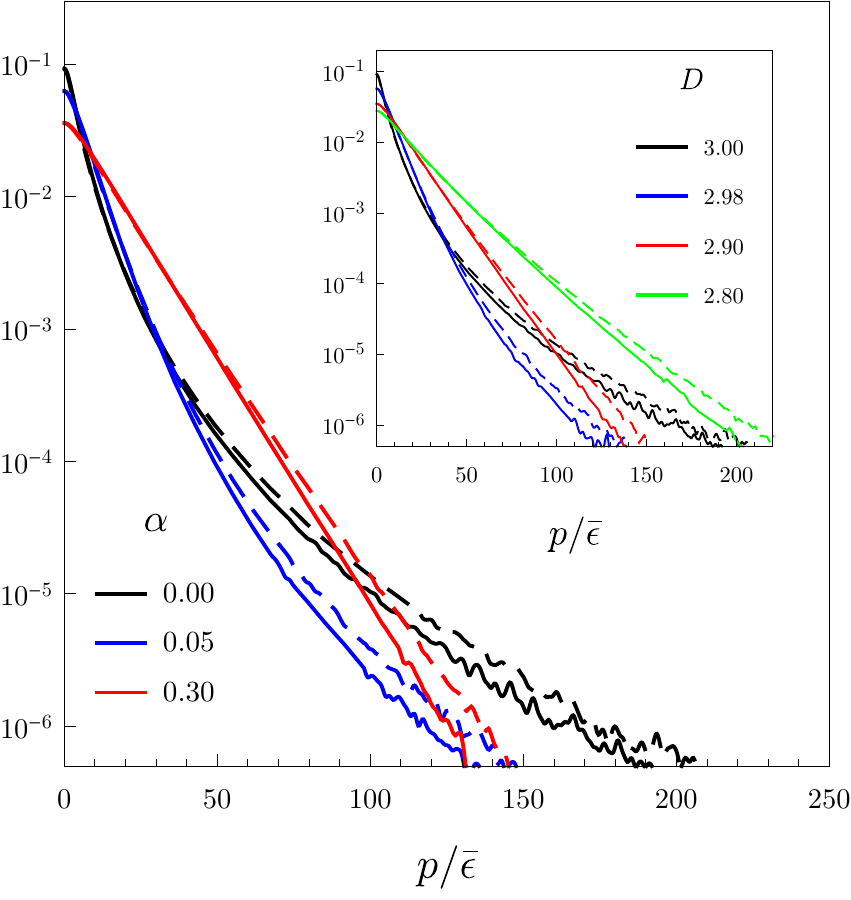}
	\caption{Pdfs of $p/\bar \varepsilon$, for homogeneously and (inset) fractally decimated NSE, along 
	with those for 3D flows.
	The negative tails, shown by broken lines, are reflected to illustrate the negative skewness of these 
	distributions. }
\label{fig:pdf}
\end{figure}

We begin our investigation by examining how the
distribution of the power $p$ is affected by the loss of intense dissipation zones caused by decimation.
In Fig.~\ref{fig:pdf} we show plots of the pdf of $p$ (with the negative tails
reflected and shown as dashed lines, for easier comparison with the positive
tails) for both non-decimated and decimated turbulence. It is visually clear
that this distribution remains negatively
skewed---energy gains are more gradual than energy losses---even in decimated flows, as a consequence of the energy cascade, despite the
suppression of extreme Eulerian dissipative regions. The tails of the distribution do become increasingly exponential, however, mirroring the transition in the shape of the pdf of $\varepsilon$ seen in Fig.~\ref{Fig:eps-pdf}.

For a more in-depth understanding, it is important to clearly identify and distinguish the contributions to the pdf of $p$, arising from trajectories passing through regions of intense dissipation, on the one hand, and mild dissipation on the other. This is especially convenient for us 
because our constant energy injection scheme allows an unambiguous measure of the mean 
dissipation $\bar\varepsilon$, and hence the conditioning of statistics on local fluctuations 
around this mean. Focusing on the non-decimated three-dimensional flow, we now calculate the conditioned pdfs of $p$ for trajectories in intense and mild dissipation zones. In practice, we identify these regions based on 
on whether the local dissipation $\varepsilon \ge 6 \bar\varepsilon$ (intense zones) or 
$\varepsilon \le \bar\varepsilon$ (mild zones). This particular choice of the upper threshold is motivated by the observation that 
the probability of $\varepsilon/\bar\varepsilon \ge 6$ is dramatically reduced even for mild levels of decimation (Fig.~\ref{Fig:eps-pdf});  
we have checked that the results that follow are insensitive to the precise choice of this threshold, in the range $4\bar\varepsilon - 6 \bar\varepsilon$.

\begin{figure}
\includegraphics[width=0.8\columnwidth]{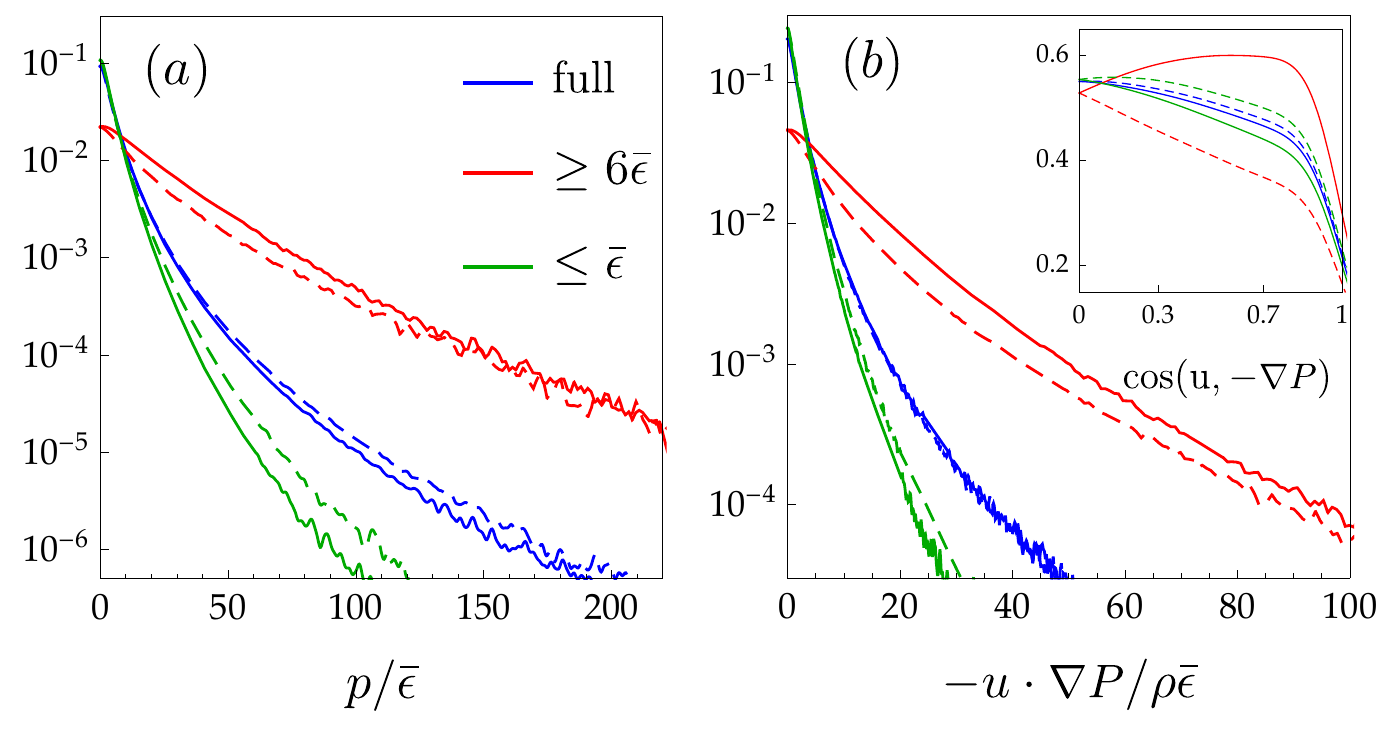}
	\caption{(a) Pdfs of $p/\bar \varepsilon$ from the full trajectory (blue) as well as from portions of the trajectory that traverse  
 regions 
	of intense (red) or mild (green) dissipation in a three-dimensional non-decimated flow.
        A comparison of the positive and negative tails (reflected and shown by broken lines) suggests a positive skewness of 
	power---\textit{take-off} events---when the trajectories sample 
	regions of intense dissipation as opposed to a negative skewness---\textit{flight-crashes}---when tracers are in 
	calmer regions. (b) The pdf of 
	$-{\bf u}\cdot\nabla P/\rho\bar\varepsilon$, as well as (inset) the alignment of ${\bf u}$ and $\nabla P$, 
	conditioned like in panel (a), showing the preferential alignment of the velocity vector with the 
	negative pressure gradient in intense regions of the flow.}
\label{fig:cond}
\end{figure}

In Fig.~\ref{fig:cond}(a), we show plots, from our non-decimated 3D simulations, of the pdf of $p$ from the full 
trajectory (blue), as well as from portions of the trajectory that traverse intense (red) and mild (green) zones. Surprisingly, 
we see that particles gain energy faster than they lose it---the opposite of flight-crashes---in 
regions of intense dissipation, because of a \textit{take-off} mechanism which we describe below. In contrast the 
flight-crash effect is more accentuated in regions of mild dissipation, thereby maintaining an overall negative skewness of the pdf of $p$.

To understand this mechanism of \textit{take-off}, we return to the
incompressible (unit-density) 3D Navier-Stokes equation, from from whence we obtain
\begin{equation}
p = -\varepsilon - {\bf u}\cdot{\nabla P}  + \nu\nabla\cdot\left[{\bf u}\cdot\left(\nabla {\bf u} + {\nabla {\bf u}}^{\rm T}\right)\right]  + {\bf f}\cdot {\bf u}.
\label{eq:power}
\end{equation} 
(We remind the reader that the term $\nu\nabla\cdot\left[{\bf u}\cdot\left(\nabla {\bf u} + {\nabla {\bf u}}^{\rm T}\right)\right]$ comes from 
the work done by viscous stresses; on averaging, this term vanishes and hence is usually not seen in energy budget equations~\cite{Davidson}.)
It is known that the leading contribution to the power comes from the
the mechanical work done by pressure gradients~\citep{flight-crash-pressure}. In regions of intense dissipation, where $\varepsilon$ is locally large, we thus have $p \approx -\varepsilon - {\bf u}\cdot{\nabla P}$. Therefore, the positive skewness of $p$ in these regions, seen in
Fig.~\ref{fig:cond}(a), can only be due to ${\bf u}\cdot{(-\nabla P)}$ being
large and preferentially positive. Evidence for this is shown in
Fig.~\ref{fig:cond}(b), which presents conditioned pdfs of $-{\bf u}\cdot{\nabla P}$.
We see that the
probability of encountering large positive values of $-{\bf u}\cdot{\nabla P}$
is indeed much higher in intense dissipation zones (red), where the pdf is strongly
positively-skewed. In contrast, the pdf shows a slight negative skewness in
mild dissipation zones (green), while it is symmetric when measured over the
entire flow domain (blue).  

To understand why $-{\bf u}\cdot{\nabla P}$ is positively-skewed in intense
dissipation zones, it is important to realize that positive values of $-{\bf
u}\cdot{\nabla P}$ arise when ${\bf u}$ is aligned with ${-\nabla P}$. This is most likely to occur when viscous forces dominate over inertial effects and balance the pressure gradient. In strongly turbulent flows, this situation is improbable except in regions where the local viscous dissipation is large.
In the inset of Fig.~\ref{fig:cond}(b), we present conditioned pdfs of
the cosine of the angle between ${\bf u}$ and ${-\nabla P}$, which confirm that
${\bf u}$ is indeed strongly aligned with ${-\nabla P}$ in zones of intense
dissipation (red). 

Lagrangian fluid particles (tracers) which encounter these intense dissipation
regions are, thus, likely to receive a strong boost of energy from the positive mechanical work
done by the local pressure gradient. This mechanical work overcomes the local
energy loss due to Eulerian dissipation, resulting in \textit{take-off}
events that give rise to the positively skewed distribution of $p$ observed in
intense dissipation zones [Fig.~\ref{fig:cond}(a)].

\begin{figure}
\includegraphics[width=0.55\linewidth]{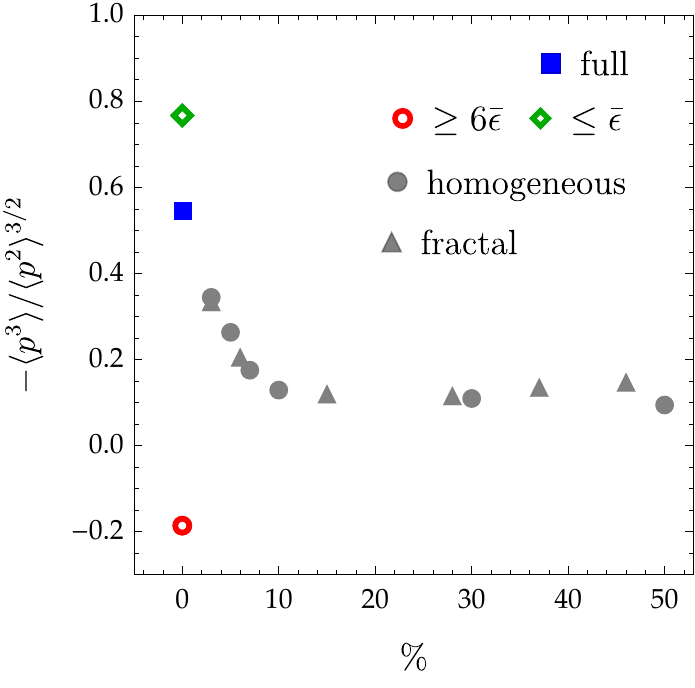}
\caption{Irreversibility Ir as a function of the degree of decimation. For the non-decimated flow, we show the irreversibility calculated using the full trajectories (Ir $> 0$, blue square), as well as that obtained from portions traversing intense 
	($Ir < 0$, red circle) or mild (Ir $> 0$, green diamond) dissipative regions. The results for the 
	decimated flows are combined by plotting Ir as a function of the percentage of modes 
	decimated~\cite{Buzzicotti-NJP}.}
\label{fig:ir}
\end{figure}

To quantify these effects, we use the third moment of the pdf of $p$ as a
measure of the Lagrangian irreversibility, and define Ir $\equiv -\frac{\langle
p^3\rangle}{\langle p^2 \rangle^{3/2}}$~\cite{Bhatnagar-Flight-Crash}. In Fig.~\ref{fig:ir},
we present Ir calculated for all trajectories in our 3D simulations, along with
the values obtained after conditioning on trajectories in zones of intense and
mild dissipation. A positive value of Ir, indicative of
flight-crashes, is obtained for mild dissipation zones (green diamond). In
stark contrast, Ir is seen to be strongly negative in intense dissipation zones
(red circle), due to the effect of pressure-gradient driven take-off events.
The overall value of Ir is positive (blue square), as the statistics are
dominated by mild-dissipation regions which occupy the majority of the flow
domain. Thus, the flight-crash behavior of tracers in turbulent flows occurs,
not because of the extreme statistics of Eulerian dissipation, but in spite of
it. 

\begin{figure}
\includegraphics[width=\textwidth]{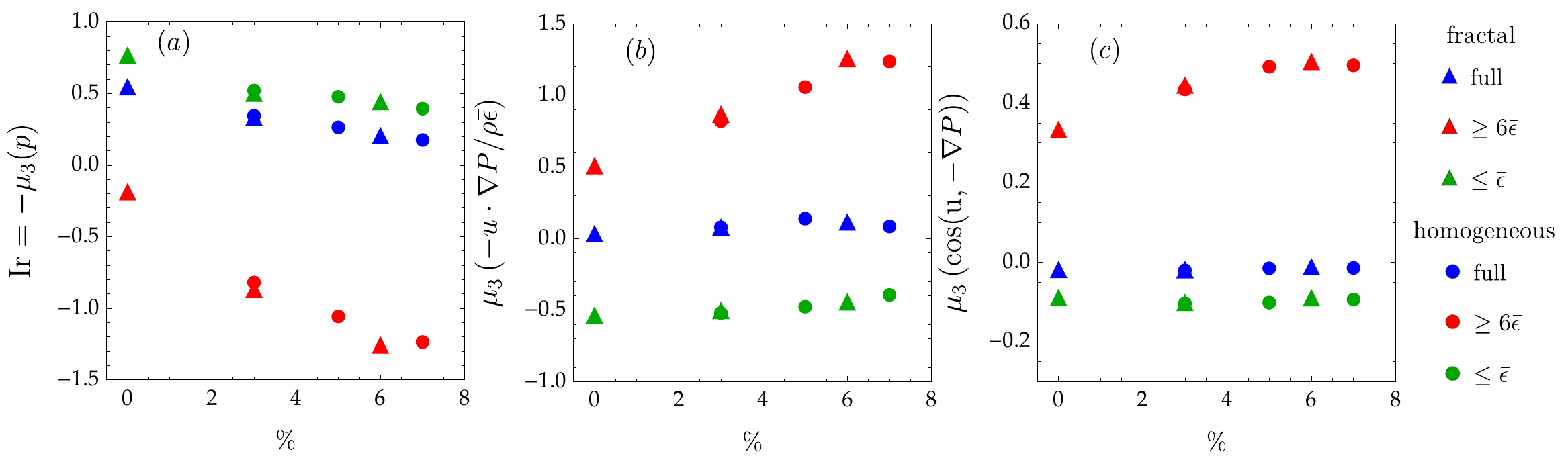}
\caption{ Influence of mild decimation on (a) Irreversibility Ir, as well as on the skewness $\mu_3(x) \equiv \langle x^3\rangle / \langle x^2\rangle^{3/2}$ of (b) the mechanical work done by pressure gradients and (c) the cosine of the angle between the velocity vector and the pressure gradient, calculated separately for regions of intense and mild dissipation, as well as for the full flow field. Both cases of fractal and homogeneous decimation are considered (see the legend), but only for small decimation levels, for which reasonable statistics on intense regions may be obtained.}
\label{fig:deci_filter}
\end{figure}

Based on this understanding, we may expect the flight-crash signature to persist even in strongly decimated flows, which are practically devoid of intense dissipation zones (cf. Fig.~\ref{Fig:eps-pdf}). This is shown to be true by Figure~\ref{fig:ir}, which presents the value of Ir for various levels of
homogeneous (gray filled circles) and fractal (gray filled triangles)
decimation. Despite an initial decrease, the value of Ir is seen to saturate
quickly to a positive value which remains relatively unchanged with increasing
decimation.

The relative contributions of intense and mild dissipation regions to the overall Lagrangian irreversibility of decimated flows is shown in Fig.~\ref{fig:deci_filter}(a). We find that intense zones continue to serve as locations for strong take-off events (${\rm Ir} < 0$), even as these zones are progressively annihilated by decimation. Indeed, this signature of take-offs appears to become more prominent in decimated flows. Furthermore, the special relationship between the velocity and pressure gradient in intense dissipation zones, which underlies the take-off mechanism, is seen to persist in decimated flows: $-{\bf
u}\cdot{\nabla P}$ has a strong positive skewness ($\mu_3$) in intense dissipation zones [Fig.~\ref{fig:deci_filter}(b)], arising from a preferential co-alignment of $-{\bf
u}$ and ${\nabla P}$ in these regions [Fig.~\ref{fig:deci_filter}(c)]. Thus, even though Eq.~\eqref{eq:power} is only applicable to non-decimated flows [because of the decimation projector $\cal P$ in Eq.~\eqref{eq:decimNS}], the intuitive understanding drawn from Eq.~\eqref{eq:power} regarding the behaviour of tracers in intense zones appears to carry over to decimated flows. Note that the results of Fig.~\ref{fig:deci_filter} are naturally restricted to mildly decimated flows, because for stronger levels of decimation the intense zones are too few to obtain good conditioned statistics.

Turbulent flows are driven-dissipative non-equilibrium systems. Therefore
their irreversibility---unlike intermittency which is an
emergent phenomenon---is not surprising, whether it be in the Eulerian or
Lagrangian frameworks, decimated or not. In this work, we uncover the
underlying correlation between the Eulerian and Lagrangian measures of
irreversibility, i.e., between the Eulerian dissipation field and Lagrangian
power-statistics. In particular, we show that regions of intense dissipation
are \textit{not} the places where tracers undergo rapid energy losses. On the
contrary, pressure gradient driven take-offs result in an inversion of the
power statistics in these intense dissipation zones.
This counter-intuitive effect is shown to result from a deceptively simple mechanism, thus adding to our understanding of the phenomenology of turbulent flows.

Our work also shows how a suppression of a small fraction of triadic
interactions leads to exponential statistics of the pdf of energy dissipation
rates instead of the familiar log-normal approximation in a 3D flow. We leave for future 
work a detailed investigation of the role of triads in the geometry and statistics 
of the Eulerian dissipation field. 

\begin{acknowledgments} 

AB acknowledges the Swedish Research Council under Grant No. 2011-542 and the
Knut and Alice Wallenberg Foundation under the project Bottlenecks for particle
growth in turbulent aerosols (Dnr. KAW 2014.0048). JRP acknowledges travel support from the  the Indo-French Centre for
Applied Mathematics (IFCAM) and the ICTS Infosys Excellence grant, as well as funding from the IITB IRCC Seed Grant. SSR acknowledges DST
(India) project ECR/2015/000361 for financial support. The simulations were
performed on the ICTS clusters {\it Mowgli}, {\it Tetris}, and {\it Mario} as
well as the work stations from the project ECR/2015/000361: {\it Goopy} and
{\it Bagha}.  

\end{acknowledgments}

\bibliography{references}

\end{document}